\documentclass[useAMS,usenatbib]{mn2e}

\usepackage{graphicx}
\usepackage{times}
\usepackage{color}

\newcommand{\CIV}{\hbox{{\rm C}{\sc \,iv}}}

\newcommand{\FeII}{\hbox{{\rm Fe}{\sc \,ii}}}
\newcommand{\SiII}{\hbox{{\rm Si}{\sc \,ii}}}

\newcommand{\Zn}{\hbox{{\rm \,Zn}}}

\newcommand{\OII}{\hbox{{\rm O}{\sc \,ii}}}

\newcommand{\MgI}{\hbox{{\rm Mg}{\sc \,i}}}
\newcommand{\MgII}{\hbox{{\rm Mg}{\sc \,ii}}}

\newcommand{\HI}{\hbox{{\rm H}{\sc \,i}}}

\newcommand{\Ly}{\hbox{{\rm Ly}$\alpha$}}
\newcommand{\Ha}{\hbox{{\rm H}$\alpha$}}
\newcommand{\Hb}{\hbox{{\rm H}$\beta$}}

\newcommand{\flux}{erg\,s$^{-1}$\,cm$^{-2}$}

\newcommand{\mpy}{M$_{\odot}$\,yr$^{-1}$}

\newcommand{\msun}{\hbox{M$_{\odot}$}}
\newcommand{\cmsq}{\hbox{cm$^{-2}$}}
\newcommand{\NHI}{\hbox{$N_{\HI}$}}

\newcommand{\LL}{\hbox{$\lambda\lambda$}}

\newcommand{\kms}{\hbox{${\rm km\,s}^{-1}$}}

\newcommand{\kpc}{\hbox{$h^{-1}$~kpc}}

\newcommand{\EW}{\hbox{$W_{\rm r}^{\lambda2796}$}}

\newcommand{\Mv}{M_{\rm h}}
\newcommand{\Ms}{\hbox{$M_{\star}$}}
       
\newcommand{\nfield}{28}		%
    
\newcommand{\fluxmin}{1.8}
\newcommand{\sfrmin}{2.9}
\newcommand{\NobsMg}{18}
\newcommand{\Nmgii}{20} 
\newcommand{\Ne}{7.6}
\newcommand{\NeEvol}{11.4}

\newcommand{\Nd}{4}
\newcommand{\PvalMgii}{0.045}
\newcommand{\PvalMgiiEvol}{0.0007}
\newcommand{\Nsigma}{2.0}
\newcommand{\NsigmaEvol}{3.4} 
 
\newcommand{\foota}{1}
\newcommand{\footc}{3} 
\newcommand{\footb}{2}

\newcommand{\HIlim}{5}
\newcommand{\NobsHI}{7}
\newcommand{\Nhi}{19}	
\newcommand{\NdHI}{2}

\def\sse{s}
\def\ssf{t}

\title[SIMPLE at $z=2$]{Enriched haloes at redshift $z=2$ with no star-formation: Implications for accretion and wind scenarios\thanks{Based on observations made with ESO Telescopes at Paranal Observatories under program ID 060.A-9041, 076.A-0527, 079.A-0341, 079.A-0600, 081.A-0568.A, 081.A-0682, 082.A-0580.}}
\author[Bouch\'e et al.]{
\parbox[t]{\textwidth}{
N. Bouch\'e$^{1,2,3}$,
M. T. Murphy$^4$,
 C. P\'eroux$^5$,
 T. Contini$^{2,3}$, 
C. L. Martin$^1$,
N. M. Forster Schreiber$^6$,
R. Genzel$^6$, 
D. Lutz$^6$,
S. Gillessen$^6$,
L. Tacconi$^6$,
R. Davies$^6$,
F. Eisenhauer$^6$
}
\vspace*{0.6cm}\\
$^1${Department of Physics, University of California, Santa Barbara, CA 93106, USA; Marie Curie Fellow}\\
$^2${CNRS; Institut de Recherche en Astrophysique et Plan\'etologie [IRAP] de Toulouse,  14 Avenue E. Belin, F-31400 Toulouse, France}\\
$^3${Universit\'e Paul Sabatier de Toulouse; UPS-OMP; IRAP; F-31400 Toulouse, France }\\
$^4${Swinburne University of Technology, Mail H30, PO Box 218, Hawthorn, Victoria 3122, Australia}\\
$^5${Observatoire d'Astrophysique de Marseille-Provence, 38 rue Fr\'ed\'eric Joliot-Curie, F-13388 Marseille, France}\\
$^6${Max Planck Institut f\" ur  extraterrestrische Physik, Giessenbachstra\ss e, D-85748 Garching, Germany}\\
}

\begin{document}

\date{Accepted 22 Jul 2011--- Received 17 Jan 2011}

\pagerange{\pageref{firstpage}--\pageref{lastpage}}  

\maketitle


\begin{abstract} 
In order to understand which process (e.g. galactic winds, cold accretion) is responsible for the cool ($T\sim10^4$~K) halo gas
around galaxies, we embarked on a program to study the star-formation properties
of galaxies selected by their \MgII\ absorption signature in quasar spectra.
Specifically, we searched for the \Ha\ line emission from  galaxies 
near very strong $z\simeq2$  \MgII\ absorbers 
(with rest-frame equivalent width $\EW\ga2$~\AA) because these
could be the sign-posts of outflows or inflows.
Surprisingly, we detect \Ha\ from  only \Nd\ hosts out of \Nmgii\ sight-lines
(and \NdHI\ out of the \Nhi\ \HI-selected sight-lines),
despite reaching a star-formation rate (SFR) sensitivity limit of \sfrmin\mpy (5$\sigma$) for a Chabrier initial mass function. 
This low success rate (\Nd/\Nmgii) is in  contrast with our $z\simeq1$  survey 
where we detected 66\%\ (14/21) of the \MgII\ hosts (down to 0.6 \mpy, 5$\sigma$).
Taking into account the difference in sensitivity  between the two surveys,
we should have been able to detect $\geq\NeEvol$ ($\geq$\Ne) of the \Nmgii\ $z\simeq2$ hosts
---assuming that SFR evolves as $\propto(1+z)^{\gamma}$ with $\gamma=2.5$ (or $\gamma=0$) respectively---
whereas we found only \Nd\ galaxies.
Interestingly,  all the $z=2$ detected hosts  have observed SFRs $\ga9$~\mpy, well above our sensitivity limit, while
at $z=1$  they all have  SFR$<9$~\mpy, an evolution that is in good agreement with the evolution of the SFR main sequence, i.e. with $\gamma=2.5$.
Moreover, we show that the $z=2$ undetected hosts are not hidden under the quasar continuum after stacking  our data.
They also cannot be outside our surveyed area as this latter option runs against our sample selection criteria ($\EW>2$\AA) 
and the  known \EW--impact parameter relation for low-ionization ions. 
Hence, strong \MgII\ absorbers
   could trace star-formation driven winds in low-mass halos  ($\Mv\leq 10^{10.6}$\msun) provided that the winds do not extend
beyond 20~kpc in order not to violate the  evolution of the absorber  number density $dN/dz(\MgII)$.   
Alternatively,
our results imply  that  $z=2$ galaxies traced by strong \MgII\ absorbers do not form stars at a rate expected (3--10\mpy) for their
(halo or stellar) masses, supporting the existence of a transition  in accretion efficiency at $M_h\simeq 10^{11}$\msun. 
This scenario can explain both the detections and the non-detections.

\end{abstract}


\section{Introduction} 

The current paradigm for  cold dark matter (CDM)  \citep{WhiteS_78a}
is now well grounded in, for instance, the many galaxy clustering surveys \citep{MadgwickD_03a,BudavariT_03a,EisensteinD_05a}.
As a result, the growth rate of dark matter haloes is well determined thanks to 
 various large-scale dark matter numerical simulations \citep[e.g.][]{SpringelV_06a}.  
For instance, in the Millennium simulation, the growth rate is determined to be 
\begin{equation}
\dot \Mv \propto \Mv{}^{\sse}(1+z)^{\ssf},\label{eq:genel}
\end{equation}
with $\ssf\sim2.2$ and a mass index $\sse$ greater than unity $\sse\simeq1.15$ \citep{BirnboimY_07a,NeisteinE_08a,GenelS_08a,McBrideM_09a}.
The index $\sse$ is greater than unity because
 the index of the initial dark matter (DM) power spectrum $n\sim0.8$ is less than unity \citep{NeisteinE_06a,BirnboimY_07a}.
Hence, this growth rate is very generic and follows from the initial dark matter (DM) power spectrum.

In contrast, our understanding of galaxy growth is more limited and is certainly incomplete.
The G-dwarf problem \citep{VandenberghS_62a,SchmidtM_63a}  and other chemical arguments \citep[e.g.][]{LarsonR_74a}
show that gas accretion must have been an important factor in building present day galaxies. 
 In addition, theoreticians keep pointing out that in haloes where the cooling time is short (compared to the dynamical time),
the accretion efficiency of `cold' ($T\sim10^4$~K) baryons must be high  \citep[e.g.][]{WhiteS_91a}. 
This mode of `cold accretion' is well defined in haloes with mass less than the shock mass $M_{\rm sh}\sim 10^{12}$\msun\
  where the cold gas is not shock-heated to the virial temperature \citep{BirnboimY_03a,KeresD_05a,KeresD_09a,vandeVoortF_11a},
and can also take place in haloes more massive than this limit at $z>2$ due to geometrical effects (streams)
\citep{OcvrikP_08a,DekelA_09a}.

While streams ---with gas columns ranging from $10^{19.5}$\cmsq\ to $10^{21.5}$\cmsq\
according to \citet{DekelA_09a}---  are expected to have small ($<5$\%) covering fractions \citep[e.g][]{FaucherG_11a,KimmT_11a},
the covering fraction of `cold accretion gas' with columns $\NHI>10^{18}$\cmsq\ can be higher ($\sim20$\%) 
in galaxies below $M_{\rm sh}$ according to \citet{StewartK_11a}.
It is therefore important to study the gaseous haloes of galaxies, which can be
 detected only in absorption in the spectra of distant background quasars (QSOs).

For instance, the low ionization \MgII\ doublet (\LL2796,2803\AA) 
is of great interest because it traces  gas at $T\sim10^4$ K
 over a wide range of redshifts $0.1\leq z\leq2.5$ and over a wide range of hydrogen column densities
$10^{16}\leq \NHI\leq 10^{22}$~\cmsq \citep{ChurchillC_00a}.
The large range of physical parameters   may thus trace the inter-stellar medium of a host \citep[e.g.][]{ProchaskaJ_97b},
parts of outflows \citep[e.g.][]{NulsenP_98a,SchayeJ_01a}, and/or $T\sim10^4$K gas related to accretion \citep[e.g.][]{KacprazakG_10a,StewartK_11a}.  

However, the sub-class of strong \MgII\ absorbers with rest-frame equivalent widths (rest-EW or \EW)  above $\sim$~1--2\AA\
are particularly interesting as they are the signposts of a collection of clouds spanning large velocity widths $\Delta v>100$\kms\
\citep{EllisonS_06a}.
At low and intermediate redshifts, the physical connection between strong \MgII\ absorbers  and star-forming galaxies 
has been established since the early nineties \citep{LanzettaK_90a,BergeronJ_91a,BergeronJ_92a,SteidelC_92a}.
Thanks to the advances in large sky surveys such as SDSS \citep[e.g.][]{AbazajianK_03a,AbazajianK_09a},
this connection is further supported
by various statistical/stacking analyses  \citep[e.g.][]{BoucheN_06c,ZibettiS_07a,MenardB_10a}. 
\citet{BoucheN_06c} showed that $z\sim0.6$ \MgII\ clouds are not virialized, a result interpreted as supporting  evidence that
supernova-driven winds are the dominant sources of strong \MgII\ with \EW$>$~2\AA~\footnote{see \citet{TinkerJ_08a} for an alternative interpretation.}.

Direct evidence for this scenario comes from our $z=1$ SINFONI Survey For Line Emitters
(z1SIMPLE) \citep{BoucheN_07a}, where we targeted two dozen QSO fields with known \MgII\ absorbers with \EW$>2$~\AA\
and successfully unveiled  the \Ha\ signature of the host in 14 out of 21 of the  cases.
These star-forming (SF) galaxies were found to have observed
 star-formation rates (SFR) of about $\sim 1$--10~\mpy, corresponding to a mean dust-corrected SFR of $\sim10$~\mpy.
This experiment further strengthens the connection between SF galaxies and strong \MgII\ absorbers with \EW\ $>2$~\AA.

Additional supporting evidence for this interpretation comes from a variety of studies.
For instance, \citet{BondN_01a} showed that,  in a few systems with \EW$>1.8$~\AA,
  the high-resolution profile of the low-ion \MgII\ shows clear signatures of a wind super-bubble.
Similarly, the study of 2 ultra-strong (with $\EW\ga$~3\AA) 
\MgII\ absorbers  of \citet{NestorD_11a} showed that 
star-burst driven outflows are necessary to account for the velocity extent of the absorption and that
each field contains a starburst (with bright [\OII], \Hb\ emission lines).
Using large data sets of 5000 and 8500 absorbers respectively, \citet{NoterdaemeP_10a} and \citet{MenardB_10a}  
found a clear correlation between \EW\ and [\OII] luminosity in the  stacked spectra
of  \MgII\ absorbers.  
These results favor outflows as the mechanism responsible for strong \MgII\ absorption, and 
as a consequence, they link intervening \MgII\ absorbers
with the blue-shifted \MgII\ absorptions seen 
in the spectra of star forming galaxies \citep[e.g.][]{MartinC_09a,WeinerB_09a,RubinK_10b}.

Given the successes of our z1SIMPLE survey and the paucity of \HI\ absorbers identified at $z\sim2$
 \citep[only 5 have been identified by][]{LowenthalJ_91a,DjorgovskiS_96a,MollerP_04a,HeinmullerJ_06a,FynboJ_10a}~\footnote{Recently,
\citet{FynboJ_11a} reported another  detection of a $z=2.58$ DLA, which we view as being tentative given its very low signal-to-noise.},
we embarked on a program designed to detect the hosts of \NobsMg\
 strong \MgII\ and \NobsHI\ \HI\ absorbers at $z\simeq2$.  
This paper presents the  results of this $z\sim2$ survey.
Throughout, we use a standard `737' cosmology, i.e. we
use a $h=0.7$, $\Omega_M=0.3$, $\Omega_\Lambda=0.7$ cosmology.

\section{Sample Selection and Data Reduction}
\label{section:data}

\subsection{Sample selection}
\label{section:sample}

The main goal of this survey is to extend the $z\sim1$ SIMPLE survey of \citet{BoucheN_07a}
to $z\simeq2$ absorbers using the $K$-band of the Spectrograph for INtegral Field Observations in the Near Infrared (SINFONI). 
 We therefore selected sight-lines  from the SDSS data base (DR5) and the 2dF Quasar Redshift survey (2QZ) data bases
 as in \citet{BoucheN_07a} with the criterion $\EW>2$~\AA.  
This criterion was used for the following two reasons.
First, in the super-wind scenario, as argued in our z1SIMPLE survey, the strongest absorbers (as measured by \EW) ought
to have the largest star-formation rates, hence the largest \Ha\ fluxes.
Second, the $\EW>2$~\AA\ criterion empirically selects the hosts with the smallest impact parameters ($\rho$)
with $\rho<35$~\kpc\ \citep[e.g.][]{SteidelC_95b}.
At $z\sim1$ and $z\sim2$, this corresponds to $\sim4$\arcsec, which means that the host-galaxy will fall within
the searched area (10\arcsec$\times10$\arcsec) limited by the SINFONI field of view (7.5\arcsec$\times$7.5\arcsec). 
From the pool of  $\sim100$  strong  $z\sim2$ \MgII\ absorbers available in SDSS, we  were able to observe \NobsMg\ \MgII\ absorbers with appropriate coordinates and favorable redshifts (see Table~\ref{table:observations}),   i.e. whose
 corresponding \Ha\ emission line would not be affected by the sky OH emission lines.

Our observations also include \NobsHI\ $z\sim2$ \HI\  absorbers (see Table~\ref{table:observations})
selected from having \HI\ column densities $\log \NHI(\cmsq)\ga20.0$ mostly from \citet{ProchaskaJ_05a}.
These \NobsHI\ SINFONI fields include 2 from archival Science Verification data (Q0216+08 and Q2243-60).  
To increase the number of \HI\ absorbers, we will include 12 other $z=2$ damped \Ly\ absorbers (DLAs)  observed with SINFONI
from the  sample of  \citet{PerouxC_11a}~\footnote{We excluded Q0405-331 which is an  associated systems with $z_{\rm abs}\simeq z_{\rm qso}$.}  and \citet{PerouxC_11c}.

Some of the targets meet both the \MgII\ and the \HI\ criteria  (SDSSJ1316, SDSSJ2059-05, Q2222-09, Q2243$-$60)
and were included a posteriori if they were not in the original sample. 
In total, our \MgII\ sample is made of \Nmgii\ sight-lines, and
the \HI\ sample is made of \Nhi\ sight-lines (see section~\ref{section:results}).

\subsection{Observational Strategies}

The observations presented here were carried out with the near-IR integral field spectrometer (IFU) SINFONI \citep{EisenhauerF_03a,BonnetH_04a}  mounted at the Cassegrain focus of the VLT UT4 telescope. 
The near-IR IFU contains a set of mirror slicers that splits
the focal plane in 32 parallel slitlets and rearranges them in a
pseudo long-slit fed into the spectrometer part of the instrument.
These reflective slicers are at the core of the high-throughput of this instrument.

In total, we have obtained $K$-band SINFONI observations towards \nfield\ $z\simeq2$   absorbing galaxies
listed in Table~\ref{table:observations} using the  $0.125$\arcsec\ pixel scale.
These observations were taken during the 2006-2009 period and 
started during SINFONI guaranteed  time observing (GTO) runs. Subsequently, the observations for this program
were obtained during the observing runs 081.A-0682 and 082.A-0580 under good observing conditions
with near-IR full-width at half-maximum  of FWHM$\simeq0.6$\arcsec.

Our $z=2$ SINFONI survey for line emitters (z2SIMPLE)
 is made of the fields selected according to the criterion described in \S~\ref{section:sample}
 (Table~\ref{table:observations}). 
Even though SINFONI is more sensitive in the $K$-band than in the $J$-band by 20\%, 
we integrated on each field for about 2hr, i.e. longer by a factor of $\sim2\times$
compared to the 40min integrations used in \citet{BoucheN_07a} in order to account for the luminosity distance
increase from $z=1$ to $z=2$. 

Despite the field of view being 7.5\arcsec$\times$7.5\arcsec, we optimized the searched area by adopting a  `on source'
dithering strategy resulting in contiguous surveyed area of 10\arcsec$\times$10\arcsec.
The central region is thus observed four times per observing block made of $4\times600$s.

\subsection{Data Reduction}

The data reduction was performed  as in \citet{ForsterSchreiberN_09a}, i.e.
 using the MPE SINFONI pipeline \citep[SPRED,][]{SchreiberJ_04a,AbuterR_06a} complemented with additional custom routines
to optimize the reduction for faint high-redshift targets such as
the OH sky line removal scheme of \citet{DaviesR_06a}. The steps are outlined here.
Firstly,  after creating the dark frames and flat-field frames,
we optimized the bad pixel identification from these frames,
as artifacts and residual bad pixels could lead to spurious sources,
and we applied these to the data.
Secondly, in order to further improve the cosmic ray removal,
we used the Laplacian edge cosmic ray removal technique of \citet{vanDokkumP_01a}. 
Bad pixels and cosmic rays were replaced
by interpolation onto the neighboring 2D pixels on the detectors.
Thirdly, arc lamp frames were used to generate the ``wavemap''  by tracing the edges and curvature of the slitlets.  
Because slight wavelength shifts may occur,
the wavemaps were   cross-calibrated against
the known vacuum wavelengths of the sky OH emission lines.
These were found in the first 600s exposure taken per field,
which was used as a reference frame.
Then, the pre-processed science data frames were reconstructed
into cubes, corrected for distortion using the tuned wavemaps. 
Given that we are looking for emission lines,
the atmospheric correction was turned off in order to avoid an additional interpolation.

Given that SINFONI is a Cassegrain instrument, small spectral shifts occur during the observations ($<0.2$pixels).
In order to ensure excellent sky-subtraction, we corrected these shifts
by cross-correlating each of the science frames spectrally against the reference frame
(the first science exposure).  
At this stage the data cubes still contain the sky background. 
We used the algorithm of
\citet{DaviesR_06a} to subtract the sky background pair-wise and optimize the OH subtraction.
The algorithm involves scaling each group of telluric OH lines separately.
We applied the heliocentric correction to the sky-subtracted frames.

For each observing block, we  use the continuum of the quasar  to spatially register the various sets of observations.  
Finally, a co-added cube is obtained from the average of all the individual
sky-subtracted 600s exposures using a median clipping at 2.5$\sigma$.

The data of the standard stars were reduced in a similar way as the science data 
and intrinsic stellar absorption lines were removed according to their spectral type.
The flux calibration of the data was performed on a night-by-night basis using the
broadband magnitudes of the standards from 2MASS. The flux calibration is accurate to
$\sim15$\%.  
Finally, the atmospheric transmission was calibrated out by dividing the science
cubes by the integrated spectrum of the telluric standard.

\section{Results}
\label{section:results}

\subsection{Low Detection Rate}

\begin{figure*}
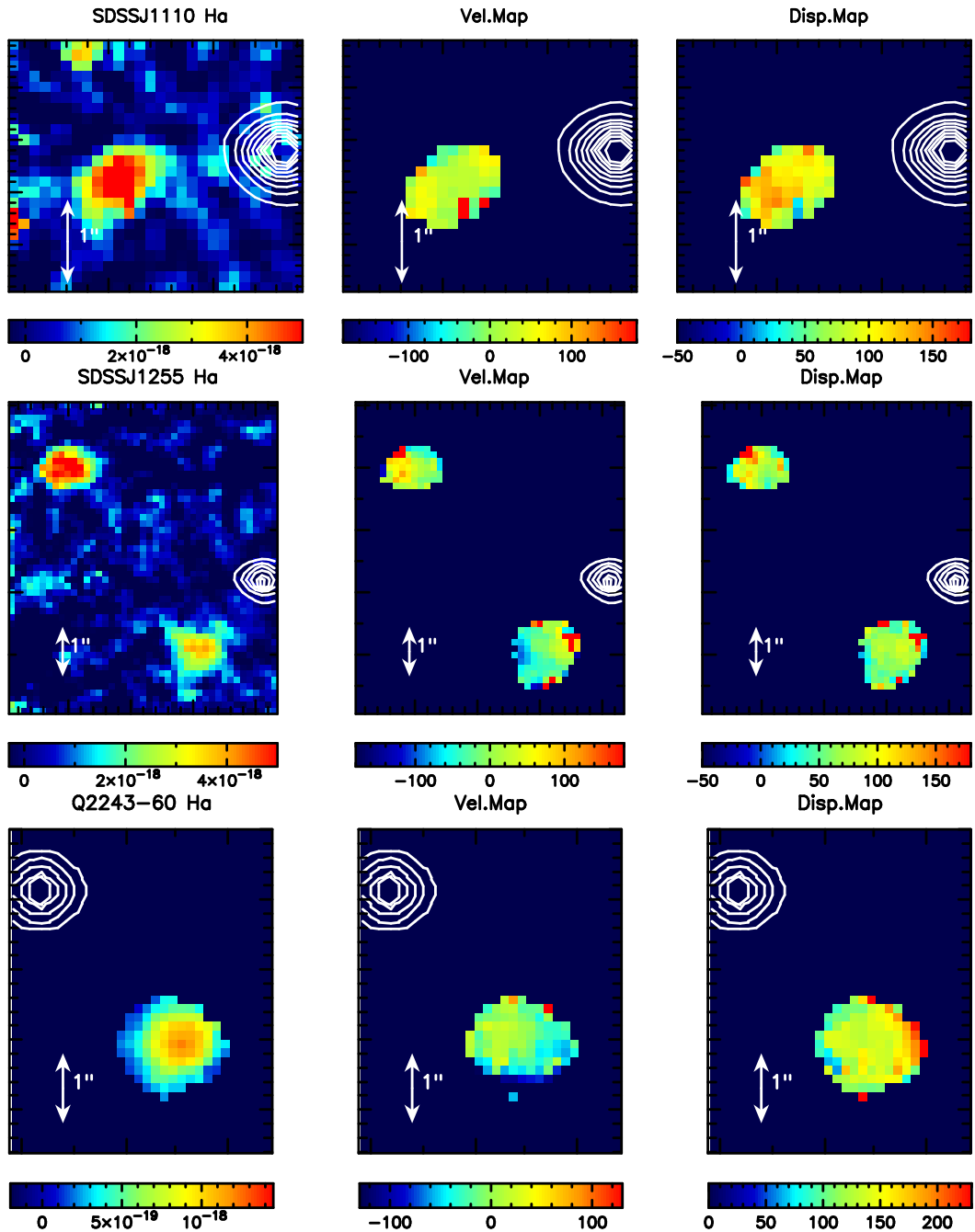
 
\centering
\includegraphics[angle=-90,width=14cm]{SDSSJ111008.61+024458.0.eps}\\
\includegraphics[angle=-90,width=14cm]{SDSSJ125525.67+030518.4.eps}\\
\includegraphics[angle=-90,width=14cm]{Q2243-60.eps}
\caption{Flux (\flux), velocity (\kms) and dispersion (\kms) maps towards SDSSJ1110$+$0244 (a), SDSSJ1255$+$0305 (b) and
Q2243$-$60 (c).  In each panel, the contours show the location of the quasar.
\label{fig:maps}}
\end{figure*}

The mean depth of our survey is determined in two independent ways. First, we measure the pixel noise 
 in a region that is within 2\arcsec\ of the QSO and at the wavelength of the expected \Ha\ line.
We   scale the noise per pixel to a region corresponding to an unresolved source of 0.8\arcsec\ (FWHM), i.e. over 32 spatial
and 8 spectral pixels to compute our $5\sigma$ flux limit~\footnote{We include  the correction for correlated noise as in \citet{ForsterSchreiberN_09a}, namely  $\sigma_{\rm real}\sim 2\times\sqrt{N_{\rm pix}} \;\sigma_{\rm pix}$.}: $\sim\fluxmin\times10^{-17}$~\flux.

Second, we inserted fake sources with known fluxes, assuming a Gaussian profile with a FWHM of 0.8\arcsec,
into our datacubes at the wavelength of the expected \Ha\ line. We attempted to detect the source and
determined the flux level where the source is no longer visible.
This technique gives similar values, i.e.   $\sim2.0\times10^{-17}$~\flux,
and is akin to a 95\%\ completeness limit.  

The flux limits for each field are listed in Table~\ref{table:summary}, and
on average, our flux limit  allows us to detect star-forming galaxies 
with SFRs greater than $>$\sfrmin~\mpy\ ---uncorrected for dust and assuming a \citet{ChabrierG_03a} initial mass function (IMF).
Despite our ability to detect star-forming galaxies with SFR$(5\sigma)\geq\sfrmin$~\mpy,
we have detected \Ha\ from very few  galaxies.
Out of our SINFONI observations towards \NobsMg\ \MgII\ absorbers, we detect \Ha\ emission of only 3 galaxies
(towards SDSSJ111008.61$+$024458.0, SDSSJ125525.67$+$030518.4 and Q2243$-$60).
Towards SDSSJ1255$+$0305,
we found 2 \Ha\ galaxies within 40\kms\ of the absorber redshift $z=2.1144$,
 and we attribute the one with the smallest impact parameter to the \MgII\ absorber.
Towards SDSSJ1144$+$0959, we detected the continuum of a galaxy 19 kpc away, however we are unable to determine its redshift.

Including the 2 sight-lines from \citet{PerouxC_11c} that meet our \MgII\ criteria a posteriori, namely SDSSJ2059$-$05 and SDSSJ2222$-$09, 
our global detection rate is \Nd/\Nmgii\ since the host towards SDSSJ2222$-$09 was also detected by Peroux et al. \citep[see also][]{FynboJ_10a}.
This detection rate (\Nd/\Nmgii) is in sharp contrast with our z1SIMPLE survey, where we   detected
two thirds (14/21) of the host galaxies \citep{BoucheN_07a}. 
We return to the significance of the low detection rate in section~\ref{section:rate}.

\subsection{Notes on the detected fields}

Figure~\ref{fig:maps}(a-c) shows the flux, velocity and dispersion maps of the 3 fields
(SDSSJ111008.61$+$024458.0, SDSSJ125525.67$+$030518.4 and Q2243$-$60, respectively) towards which we could detect
the host. 
The host of the absorber towards SDSSJ111008.61$+$024458.0 is found only 17kpc North from the QSO.
The absorber at $z=2.1187$ has $\EW=2.97\pm0.17$\AA\ \citep{MurphyM_04c,ProchaskaJ_05a}
and is found to have a SFR$=9.1$\mpy\ for a Chabrier IMF, corresponding to a dust-corrected SFR$_0$ of $\sim19$\mpy.
Its velocity field is indicative of a merger system with a reversal of the velocity field, hence we refrain ourselves in
assigning a dynamical mass.
The average dispersion of $\sigma\sim120$~\kms.

The host of the absorber towards SDSSJ125525.67$+$030518.4 with $\EW=3.20\pm0.23$\AA\ \citep{MurphyM_04c,ProchaskaJ_05a}
is found at 14kpc from the QSO.
The host has a SFR$=8.6$\mpy\ calculated also assuming a Chabrier IMF, corresponding to a dust-corrected SFR$_0$ of $\sim18$\mpy.
The velocity field shows a clear rotation pattern of $v_{\rm max}\sin i\sim\pm70$\kms\ with a  dispersion that peaks at $\sigma\sim70$~\kms\ in the center.
The intrinsic dispersion is $\sigma_0\sim30$\kms\ in the outskirts.  The dynamical mass within the 
 half-light radius (0.44\arcsec) is $ M_{\rm 1/2}\sim2.0\times10^{10}$\msun, taking into account
the inclination $i=24^{\circ}$ estimated from the axis ratio $b/a=0.9$.
Using $v_{\rm max}$, our estimate of the haloe mass is $\Mv\sim2.5\times10^{11}$~\msun.

The host towards Q2243$-$60 has the highest \Ha\ flux in our sample, reaching almost $\sim 1\times10^{-16}$\flux, corresponding to an
 observed~\footnote{Using $\Ha/Hb$, we find, in Bouch\'e et al. (in preparation),
 that $E(B-V)\sim0.6$ and therefore the dust-corrected SFR is SFR$_0$ of $\sim77$\mpy.} SFR$=18$\mpy.   
The velocity field also shows a rotation pattern with a large central
dispersion $\sim120$\kms.  This galaxy is less resolved (its half-light radius $\sim0.5$\arcsec\ is similar to the PSF$\sim0.6$\arcsec).
We recently obtained   SINFONI observations (in $H$ and $K$) with adaptive optics (AO) for this field that will be presented elsewhere
 (Bouch\'e et al. in preparation).  
We can already report that the velocity field is resolved with $v_{\rm max}\sin i\simeq 100$\kms, the
intrinsic dispersion is $\sigma_0\sim100$\kms.
With an inclination of $57\pm2$~$^{\circ}$,  the dynamical mass within the half-light radius (0.45\arcsec) is $ M_{\rm 1/2}\sim1.8\times10^{10}$\msun.
The halo mass is estimated at $\Mv\sim2\times10^{11}$~\msun\ from $v_{\rm max}$.

This absorber has a large \MgII\ rest-frame equivalent width $\EW=2.6$~\AA\ corresponding to a velocity width of $\sim250$\kms\ \citep{EllisonS_06a}. 
 This DLA was discovered by the Hamburg/ESO QSO Survey  \citep{ReimersD_97a}.
Its \HI\ column density is  $\log [ \NHI/\cmsq] = 20.7$. 
\citet{LopezS_02a} presented a deep (28ksec) UVES spectra of this QSO with S/N ranging from $>50$ to 80.
These observations show that this DLA has an absorption metallicity that is [Zn/H]$=-1.10\pm0.05$, close
to the iron abundance  [Fe/H] $=-1.26\pm0.02$, indicating a low dust depletion.

\subsection{How many detections are expected?}
\label{section:rate}

Before interpreting this apparent low detection rate (\Nd/\Nmgii), we need  to be certain  
(i) that the sample properties are similar
and (ii) that  our $z=1$ sources could have been detected at $z=2$. 
On the first point, a KS-test for the $\EW$ distributions gives a probability of $P=0.93$,   
showing that the two distributions are very consistent with each other.
In other words, the low success rate cannot be attributed to differences
in our sample selection.  We also found no differences in the doublet ratio (DR$=W_r^{2803}/W_r^{2796}$), nor in \MgI\
rest-EW distributions.  However, we note that the parent $z=2$ sample made of 75 absorbers with $\EW\ga2$\AA\
has a \FeII2600\AA/\MgII2796\AA\ ratio that is lower than the parent $z=1$ sample by 30\%\ (0.4 vs. 0.6).

We now  quantify   whether the galaxies in our z1SIMPLE survey
 could have been detected at $z=2$.
While the answer to this question depends directly on the flux limit of our survey,
it also depends on the assumed evolution of the SFR properties from $z=1$ to $z=2$.

For the sensitivity aspect, our z2SIMPLE survey is more sensitive  
($\sim\fluxmin\times10^{-17}$~\flux) than our z1SIMPLE survey ($2.6\times 10^{-17}$~\flux\  computed over the same `volume' of pixels),
thanks to the increased throughput in $K$ and longer exposure times.
In spite of better flux limits, 
our $z=2$ SFR limit ---computed for a Chabrier IMF--- is \sfrmin~\mpy\ (not corrected for dust), i.e. $\sim 5\times$ worse than at $z=1$
(0.6~\mpy, not corrected for dust)~\footnote{Note in \citet{BoucheN_07a},
we quoted dust-corrected  3--$\sigma$ limits.}
given the increase in luminosity distance ($D_L\propto(1+z)^2$).
See Table~\ref{table:summary} for the individual limits taking into account the redshift  dependent noise.

\begin{figure}
\centering
\includegraphics[width=9cm]{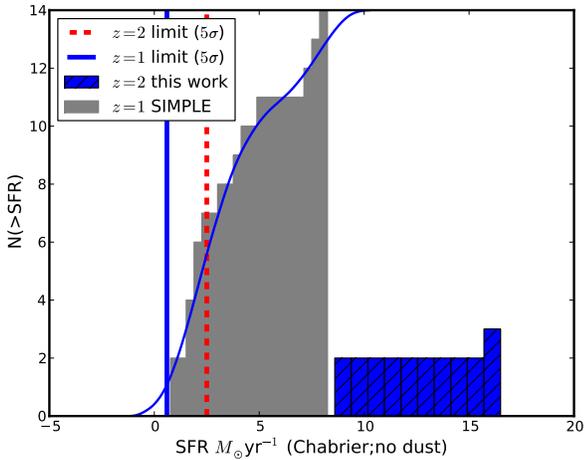}
\caption{Cumulative observed SFR distributions.  The curve shows the smoothed distribution.
The grey (hatched) histogram shows the $z=1$ ($z=2$) distribution, respectively.
At high-redshifts, all the detections have SFR$\ga9$~\mpy, while none have such high SFRs in the $z=1$ survey.
The detection limits are indicated as vertical lines and show that at most 6 of the 14 $z=1$ galaxies would not
have been detected in this survey (assuming no evolution). }
\label{fig:SFRdistri}
\end{figure}

From the cumulative SFR distribution of the $z=1$ SIMPLE survey \citep{BoucheN_07a}
 shown in Figure~\ref{fig:SFRdistri}, 
we see that 6 of the 14 $z=1$ detected galaxies would not have been detected at $z=2$ given our survey completeness limit (vertical dashed line). 
Similarly, using 10,000 Monte Carlo resamplings, we find that 99\% of the time 6 galaxies would not be detected
 if the SFR distributions were the same at both epochs. 
However, the SFR distribution for the detected hosts is   very different  at the two epochs
as   Figure~\ref{fig:SFRdistri} shows.
At $z=2$, all the detections have SFR$\ga9$~\mpy, while none have such high SFRs in the $z=1$ survey.
Thus, if the SFRs evolve strongly, as suggested by the three detected hosts in Figure~\ref{fig:SFRdistri}
and by the evolution of the SFR--Mass sequence, which evolves as $\propto(1+z)^{2.5}$ (see Eq.~\ref{eq:sfrmass} below),
then only 2 of the 14 $z=1$ hosts would not have been detected. 
Therefore,  we expect a detection or `success' rate of $\hat p=0.38$ (8/21), assuming no evolution of the SFR distribution,
and of $\hat p=0.57$ (12/21) assuming a SFR evolution going as $\propto(1+z)^{2.5}$.

Hence, we ought to find $\geq\Ne$ or $\geq\NeEvol$ hosts depending on the assumed SFR$(z)$ evolution, whereas  \Nd\ galaxies were found.
Treating the  individual fields as independent experiments with a success rate given by $p$, the number of success
follows a binomial distribution~\citep{CameronE_10a}.  
The probability to have only \Nd\ successes from such a binomial
distribution is \PvalMgii (\PvalMgiiEvol), i.e.  the $z=2$ detection rate is different than the $z=1$ detection rate at \Nsigma$\sigma$
or \NsigmaEvol$\sigma$ assuming SFR($z$) goes as $\propto(1+z)^{\gamma}$ with $\gamma=0$ or $\gamma=2.5$, respectively
(see table~\ref{table:results}). 

\subsection{Closing loopholes}

Before interpreting these results, we ask the following questions:
(i) can the hosts be outside the SINFONI field of view?
(ii) can the \Ha\ host emission be hidden under the QSO continuum?

{(A) Are the $z=2$ hosts outside the field of view?}
By design, both the $z=1$ and this $z=2$ sample were selected to have rest-\EW$>2$~\AA.
This criteria corresponds to a maximum impact parameter $\rho_{\rm max}\sim35$~kpc given the well-known
anti-correlation between \EW\ and impact parameter \citep[e.g.][]{LanzettaK_90a,SteidelC_95a,ChenHW_10a} for $z\sim1$ systems
and was chosen to ensure  that the host would fall within our mapping region ($\sim10\arcsec\times10$\arcsec\ or 80~kpc$\times80$~kpc).
The normalization of the \EW--$\rho$ anti-correlation may expand between the two epochs. 
However, as the Universe was denser
at $z=2$, it is difficult to imagine that that the normalization actually expanded.
Not only this would be opposite to the galaxy and  halo size evolution, but it would also
 contradict the observations of \citet{ChenHW_10a} and  \citet{SteidelC_10a}. 
\citet{ChenHW_10a} showed that the scatter in the \MgII\ EW--$\rho$  anti-correlation scales with stellar mass $\propto M_{\star}^{0.28}$
(and weakly on sSFR). In other words, our results would imply the opposite, namely that the scatter scales inversely proportional to $M_\star$.
Similarly,  \citet{SteidelC_10a} showed that the averaged
rest-EW of another low-ion (\SiII\ with similar ionization potential as \MgII) is also anti-correlated with impact parameter.
They found that the (average) rest-EW of \SiII1526 is much less than $\la0.3$\AA\ at impact parameters $\ga$40~kpc.  
We have stacked our SDSS spectra and found that on average the non-detections have a rest-EW $\sim0.8$~\AA\ for \SiII1526.
In other words, in conjunction with the \citet{SteidelC_10a} results, our sample 
of $>2$\AA\ \MgII\ absorbers has no properties consistent with being outside our mapping region.

{(B) Are the $z=2$ hosts too close to the QSO?}
An IFU is a very good tool to untangle \Ha\ emission even with impact parameters
less than the QSO point-spread-function (PSF). This is nicely demonstrated by the J0226$-$28 sight-line
in our $z=1$ sample \citep{BoucheN_07a} where the \Ha\ emission is detected 0.25\arcsec\ from the QSO PSF (FWHM$\sim0.8$\arcsec).  

\begin{figure}
\centering
\includegraphics[width=10cm]{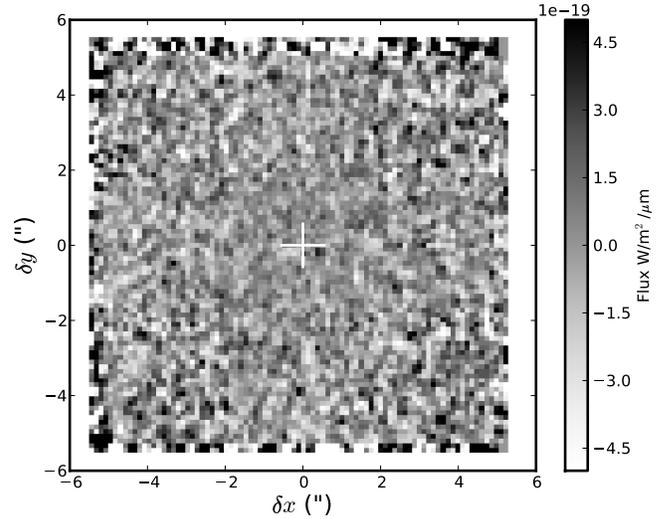}
\caption{Continuum subtracted image of the stacked SINFONI cube extracted around \Ha\ of the 16 non-detected
absorbers. The cross shows the location of the (stacked) continuum.
 We find no evidence for \Ha\ emission in the stacked cube down to $\sim10^{17}$\flux, i.e.  SFR~1.5$\sim$,
ruling out the possibility that the hosts were within 0.3\arcsec\ from the QSO. \label{fig:stacked}}
\end{figure}

In the present survey, however, it is possible that the \Ha\ emission is hidden under the QSO continuum,
i.e. within a radius of 0.3\arcsec\ from the QSO.
In order to address this, we dereshifted  and stacked the 16 SINFONI cubes of the non-detected subsample.
The continuum-subtracted image, shown in Fig.\ref{fig:stacked}, reaches
the same flux limits within 0.3\arcsec\ of the QSO PSF as the ones quoted in section~3.1. 
Therefore, we can say with confidence that we find no evidence for \Ha\ emission in the stacked SINFONI cube,
ruling out the possibility that the hosts were too close to the QSO.

\section{Implications for the nature of \MgII\ systems.}
\label{section:implication:mgii}

Having established that we are not limited by the QSO continuum, nor by the field of view,
 the low detection rate show clearly that, in the most neutral terms, 
the $z=2$ \MgII\ population probes gaseous haloes that are different than at $z=1$.
We now attempt to constrain the meaning of `different'
by placing  limits on the host mass assuming known scaling relations and the evolution of $dN/dz$.

\subsection{Mass constraints from the SFR sequence.}
\label{section:massconstraints}

In the mass regime where cooling times are short (i.e. where cold accretion dominates),
  SFR is driven by the (net) baryonic accretion rate as shown in \citet{DekelA_09a,DuttonA_10a} and \citet{BoucheN_10a}.
The baryonic accretion rate is then simply given by   the cosmological baryonic fraction $f_B$ times
the dark-matter accretion rate $\dot M_h$ (Eq.~\ref{eq:genel}) which is set by the cosmological parameters \citep[e.g.][]{NeisteinE_06a}.
Hence, for a given accretion efficiency $\epsilon$, we have 
\begin{equation}
\hbox{SFR}\sim \epsilon\;f_B\;\dot{\Mv}\sim 6 \;\epsilon_{1.0} M^{1.15}_{h,12}(1+z)^{2.25} M_{\odot}~\hbox{yr$^{-1}$}, \label{eq:accr}
\end{equation}
where $M_{h,12}$ the halo mass in units of  $\Mv/10^{12}\msun$
and we use $\epsilon_{1.0}\equiv\epsilon/1.0$  since
\citet{GenelS_08a} \citep[see also][]{DekelA_09a,BoucheN_10a} showed that the accretion efficiency $\epsilon$ must be in the 70-80\%\ range
in order to account for the large SFRs of $z=2$ star-forming galaxies~\footnote{Our assumption of $\epsilon=1.0$
is not -- as it may seem -- extreme. Indeed, a lower and more realistic efficiency would be compensated by the more appropriate net SFR,
 which is $\sim1/2\times$SFR  if SN outflow rates are proportional to SFRs. Hence, $1/2$SFR$\sim \epsilon_{0.5} f_b\dot M_h$, which is equivalent to equation~\ref{eq:accr}.}.
This equation allows us to put limits on the halo mass of the hosts, but before applying it to our survey,
we demonstrate its validity on several galaxy samples.

The expected SFRs determined from Eq.~\ref{eq:accr}  
agree well with the observed SFRs at a given halo mass for a wide variety of galaxies.
For instance, Lyman break galaxies (LBGs), which have halo masses $\sim10^{11.8}$\msun\ from their clustering
\citep{AdelbergerK_05a} or from their rotation curves \citep{ForsterSchreiberN_06a}, have an expected SFR
in the range of 30--100\mpy, which is in good agreement with the observed values for $z=2$ star-forming
galaxies \citep[e.g.][]{ForsterSchreiberN_09a,ErbD_06c}.
Similarly, the $z=1$ \MgII\ absorbers with $\EW>2$~\AA\  reside in haloes with $\Mv\sim\times10^{11.3\pm0.4}$~\msun,
as several clustering analyses 
\citep[e.g.][]{BoucheN_06c,GauthierJR_09a,LundgrenB_09a} 
have shown, and are expected to have SFRs in the range few--10\mpy, 
in very good agreement with the observed SFR distribution shown in Fig.~\ref{fig:SFRdistri}.

Now that we have validated  Eq.~\ref{eq:accr}, we turn it around using our SFR upper limit to put limits on the halo mass
of  our $z=2$ non-detected sample.  Our limit of SFR$<\sfrmin$ \mpy\
 corresponds to low mass haloes with masses $\Mv\la 4\times 10^{10}$~\msun\ ($V_h\sim$65\kms).
Similarly, the SFR--\Ms\ relationship \citep[e.g.][]{ElbazD_07a,DaddiE_07a,DroryN_08a,SantiniP_09a,PannellaM_09a,DamenM_09a,KarimA_11a}
can be used to place a limit on the stellar mass $\Ms$. These results show that the SFR--\Ms\ sequence is
\begin{equation}
\hbox{SFR}\sim 150 \Ms_{11}^{0.8} (1+z)^{2.7}_{3.2},\label{eq:sfrmass}
\end{equation}
where $\Ms_{11}\equiv\Ms/10^{11}\msun$ and $(1+z)_{3.2}\equiv(1+z)/3.2$.  Our limit of SFR$\la$\sfrmin~\mpy\
 corresponds to $\Ms\simeq6\times10^8$~\msun\ if we extrapolated the SFR--sequence to low masses.
Note, this stellar mass agrees well with the limit on the halo mass we just placed and is thus not an independent result.
Indeed, these numbers would place such a host along the low-mass end of the known $\Ms$-$\Mv$ relation 
\citep{ShankarF_06a,vandenBoschF_07a,GuoQ_10a,MosterB_10a,BehrooziP_10a}.

In summary, the host of strong \MgII\ absorbers could in principle be in very low mass haloes with $\Mv\la 4\times 10^{10}$~\msun,
i.e. below our SFR limit~\footnote{Note that despite having a narrow range of \EW, the range of halo mass could be larger. 
Such a scatter is enough to explain our few detections.}.
In the next section, we ask whether this   low-mass conclusion is  consistent with the observed evolution of $dN/dz$.

\subsection{Constraints from $dN/dz$.}

The number of absorbers per unit redshift $dN/dz$ is a well measured quantity  \citep[e.g.][]{NestorD_05a,ProchterG_06a}.
A more appropriate quantity in our context is the number of absorbers per unit co-moving distance $dN/dX$
and these surveys have shown that $dN/dX$ does not evolve from  $z=1$ and $z=2$.
The number of absorbers per unit co-moving distance, $dN/dX$, for $z=1$ absorbers  is:
\begin{eqnarray}
\frac{{\rm d}N}{{\rm d}X}(>2\AA; z=1)&\equiv&0.02\frac{n_{11.3}}{10^{-2}}\left (\frac{R}{35\hbox{kpc}}\right )^2,\label{eq:dn}
\end{eqnarray}
where $n$ is the halo number density, $\sigma=\pi R^2$ is the physical cross-section
and $n_{11.3}\equiv \Mv/10^{11.3}$ in Eq.~\ref{eq:dn}.

Because $n(M_h,z)$, the co-moving density of  haloes, is a very weak function of $z$
from $z=0$ up to $z=5$ \citep{MoH_02a}, we can say that
any evolution of the absorber cross-section will imply an evolution of the typical halo mass $M_h$ via:
\begin{eqnarray}
\frac{{\rm d}N}{{\rm d}X}(>2\AA; z)&\equiv&0.02\frac{n_{11.3}}{10^{-2}}\left (\frac{R}{35\hbox{kpc}}\right )^2\left (\frac{1+z}{1.8}\right)^{-2m},\label{eq:dNdz}
\end{eqnarray}
for any index $m$ of the evolution of the cross-section.
The physical cross-section of galaxies (and haloes) are indeed smaller at higher redshifts as demonstrated by the numerous groups
\citep[e.g.][]{DahlenT_07a,WilliamsR_10a,MoslehM_11a}. These surveys indicate that the size evolution index $m$ in Eq.~\ref{eq:dNdz} is $\sim1$.

For such an index $m=1$, the redshift factor in Eq.~\ref{eq:dNdz} is then about 0.3$=\left[({1+2.2})/({1+0.8})\right]^{-2}$,
which means that the density of halo $n(\Mv)$ must be higher by
a factor 3$\times$ in order to account for the constant dN/dX.
In the previous section, we have established that the hosts of our \MgII\ sample might be in very low-mass galaxies,
with halo masses $\Mv\la 4\times 10^{10}$~\msun.  Such low-mass haloes are more numerous
by a factor of $\sim 4$ \citep[e.g.][]{MoH_02a}.   
Therefore, the cross-section evolution could compensate for the higher number density of low-mass haloes
provided that the cross-section evolves.

\subsection{Implications for the nature of \MgII\ absorbers}

In order to account for our low success rate, in the last section we discussed two possibilities, namely the $z=2$ hosts
could reside (i) in haloes with similar masses to those at $z=1$ ($\Mv\sim10^{11.3}$ \msun), or
(ii) in haloes with very low masses  ($\Mv\sim10^{10.6}$ \msun) below our flux limit.
The first option violates our observations because  such haloes are expected to have  SFRs $>10$ \mpy\ from
the SFR--\Ms\ or SFR--$M_h$ sequence, 
whereas we find only 3 such galaxies down to \sfrmin\mpy.

\subsubsection{Wind scenario}

The second option is equivalent to a scenario invoking winds in low mass galaxies.
In such a scenario, 
the large equivalent widths of \MgII\ are the signature of `cold' gas ($T\sim10^4$~K) entrained in outflows at a few hundreds of \kms.
This option would explain our low detection rate
but --as we showed-- it is not consistent with the observed
number density of absorbers,  $dN/dX$, {\it unless} the cross-section of absorbers has evolved strongly as $\propto(1+z)^{-1}$. 
Given that the $z=1$ cross-section of strong \MgII\ absorbers is typically $\rho\sim35$--40~kpc \citep[e.g.][]{SteidelC_95a,BoucheN_06c}, 
the cross-section of $z=2$ absorbers cannot then be larger than 20~kpc.

Hence, our results are consistent with the outflow scenario as argued by others \citep{BondN_01a,BoucheN_06c,BoucheN_07a,NestorD_11a},
provided that the $z=2$ hosts reside in halos less massive than $\Mv\sim10^{10.6}$ \msun\ {\it and}
that the extent of the wind traced by $\EW>2$\AA\ is not larger than 20~kpc ($\sim$~2.5\arcsec).
A consequence of this scenario is that the outflow `efficiency'  (defined as \EW/SFR, i.e.  more clouds (larger \EW) per unit SFR  are being produced at $z=2$ than at $z=1$) was much larger at $z=2$ than at $z=1$.
However, this scenario cannot explain the few galaxies that we do detect. Indeed, because
the SFRs are higher at $z=2$ than at $z=1$, this $\EW$-to-SFR ratio is then lower at $z=2$, not higher.
Thus, the wind scenario alone cannot explain the detections and non-detections at the same time.

\begin{figure}
\centering
\includegraphics[width=8cm]{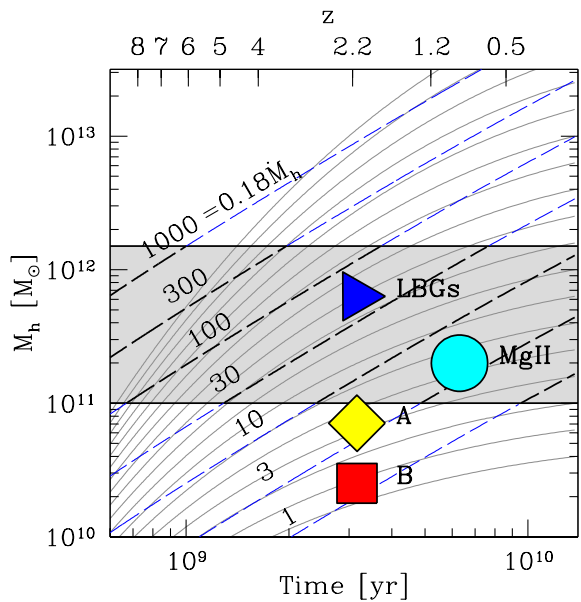}
\caption{Each solid curve represents a halo growth history $M_h(z)$.
The cosmological (baryonic) growth rate, namely $\dot M\epsilon\; f_B\;\dot M_h$
where $f_B$ is the baryonic fraction, is shown by the dashed lines.
These dashed lines can be used as a proxy for SFR when (cold) accretion is efficient with $\epsilon\simeq1.0$  \citep[e.g.][]{GenelS_08a},
i.e.  below the shock mass $M_{\rm sh}10^{12}$\msun, and possibly only in the grey band
according to \citet{BoucheN_10a} and \citet{CantalupoS_10a}.
For illustration purposes, the inferred SFRs for LBGs (triangle) and strong $z\sim1$ \MgII\ absorbers (circle) 
are  in very good agreement with the observed SFRs (see text).
Strong $z=2$ \MgII\ absorbers could be caused  either in galaxies with low accretion efficiency (scenario A)
or in low mass SF galaxies below our SFR limit of $\sfrmin$\mpy\ (wind scenario B) provided that their cross-section
is $<20$~kpc in order to match the observed evolution of $dN/dz$.
}\label{fig:MyFig}
\end{figure}

In Fig.~\ref{fig:MyFig}, we put our $z=2$ results in a more global context.
The solid lines show the averaged evolution of halos with time $M_h(z)$ and the dashed lines
show the predicted SFR ($\sim\epsilon f_B \dot M_h$) given by Eq.~\ref{eq:accr} for a maximum accretion efficiency $\epsilon=1.0$.
Samples with known halo masses such as LBGs \citep{AdelbergerK_05a} and $z=1$ \MgII\ absorbers \citep{BoucheN_06c} are shown by the large triangle and
circle respectively.  As mentioned in section~\ref{section:massconstraints}, 
the expected SFRs determined by Eq.~\ref{eq:accr} for these  samples 
agree well with the observed SFRs leading to the conclusion the accretion efficiency $\epsilon$ must be high \citep[e.g.][]{GenelS_08a}.
The wind scenario in low mass galaxies is represented by the large filled square (labeled B) in Fig.~\ref{fig:MyFig}.

\subsubsection{Accretion scenario}

A third possibility is that  \MgII\ absorbers  reside in haloes of intermediate mass with $\Mv\sim10^{10.6\hbox{--}11.0}$\msun,
i.e. between the two extremes of $10^{10.4}$\msun\ and $10^{11.3}$\msun\ already discussed.
 This mass scale for the $z\sim2$ hosts is illustrated by the large diamond (labeled {\bf A}) in Fig.~\ref{fig:MyFig}.
This mass scale falls {\it below} the grey band in Fig.~\ref{fig:MyFig}. 
The grey band illustrates the regime where the accretion efficiency (or cooling efficiency)  is certain to be high.
This band is defined by a low and high mass.
The high mass represents the shock mass $M_{\rm sh}\sim10^{12}$\msun, the well-known transition between cold and hot accretion \citep[e.g.][]{BirnboimY_03a,KeresD_05a,DekelA_06a,KeresD_09a,DekelA_09a}.
The  lower mass $M_{\rm min}\sim10^{11}$\msun\  represents
another possible transition~\footnote{Note that the power law $\propto \Mv^{\beta}$ index at this transition
is directly related to the slope of the luminosity function  \citep[see also][]{KravtsovA_09a}.}
 in the cooling properties of haloes \citep[e.g.][]{CantalupoS_10a}.
This $M_{\rm min}$ hypothesis  was postulated in  \citet{BoucheN_10a} because these authors realized a simple analytical model
with this single hypothesis could reproduce several key observational results, such as the  Tully-Fisher relation, the SFR--\Ms\ sequence, the downsizing phenomenon, the gas fractions ($f_g$), stellar fractions, and the cosmic star-formation history ($\dot \rho_{\star}$).

Since we did not detect the majority of the galaxies, 
this intermediate mass scale scenario (labeled A in Fig.~\ref{fig:MyFig}) implies that the efficiency $\epsilon$ for accretion
in Eq.~\ref{eq:accr} must be  much smaller than our fiducial value of $\epsilon=1.0$,
otherwise the SFRs would have been   above our observational limit.  Turning it around,
the accretion efficiency $\epsilon$ should have evolved by about $2.5\times$ from $10^{10.8}$~\msun\ to $10^{11.3}$\msun,
i.e. this scenario would imply that $\epsilon(\Mv)$ goes as $\Mv^{\beta}$ with $\beta>1$.
In other words, the mass dependence of $\epsilon$ is super-linear. This
has the advantage that this scenario can explain both the detections and the non-detections by invoking
a finite scatter around a $\Mv$--\EW\ relation.

\section{On the nature of \HI-selected absorbers.}

As for the \MgII\ sample, the only criterion used when selecting \HI\ absorbers is the redshift
 to ensure that the spectral region expected for \Ha\ in SINFONI
is free of bright sky emission lines. From the samples available in the literature,
we observed \NobsHI\ fields, and  detected  only one host  (see Table~\ref{table:summary}).
We also include the 12 $z=2$ sight-lines from \citet{PerouxC_11c}, leading to a total sample of \Nhi\ \HI-selected absorbers.
Admittedly, this \HI-selected sample does not cover the whole range of \HI\ columns and metallicities for DLAs,
but its mean column density is that of a typical DLA.

In all, out of \Nhi\ \HI-selected  sight-lines, only \NdHI\ $z\simeq2$ hosts have been detected so far with SINFONI.   
Contrary to the \MgII\ systems, we cannot give a statistical significance to this result since the $z=1$ success rate for \HI-selected absorbers is poorly constrained.  
We note that the  $z=1$ success rate for \HI-absorbers appears somewhat higher:
out of 9 $z=1$ sight-lines observed with SINFONI,   \citet{PerouxC_11a} and \citet{PerouxC_11c}
detected the host towards  4 \HI-absorbers.

Thus, our main result for \HI-selected absorbers is similar to that of our \MgII-selected sample and to
many other $z\sim2$ DLA surveys, which have also failed to unveil large numbers of hosts using a range of techniques. 
This is a surprising outcome given that the effort invested and given that sub-DLA/DLAs cover almost the entire sky \citep[e.g.][]{PerouxC_03a,ProchaskaJ_09a,NoterdaemeP_09a}. 
Among the most sensitive recent studies,  
there is the stacking analysis of 341 DLAs with $\log \NHI(\cmsq)\ga20.62$ by  \citet{RahmaniH_10a} who could only place a $3\sigma$ upper limit of $3.0\times10^{−18}$~\flux\ on the average \Ly\ flux, corresponding to 0.03$L^*$(\Ly). 
There is also the analysis of \citet{WolfeA_06a} who searched for in situ star formation in DLAs by looking for low surface brightness emission in the Hubble Ultra Deep Field (HUDF) F606W image and could also only place upper limits on the SFRs surface density. 
These upper limits were, surprisingly, at least a factor 10 lower
than the rates predicted by the distribution of neutral-gas column
densities in DLAs, assuming a normal SF law.

A complete review of recent literature results is beyond the scope of this paper, but in general very few DLA hosts have been identified
in spite of numerous search campaigns during the past two decades \citep[e.g.][among others]{DeharvengJ_90a,LowenthalJ_90a,LowenthalJ_95a,BunkerA_95a,BunkerA_99a,MollerP_02a,KulkarniV_00a,KulkarniV_06a}.
Indeed, these efforts led to only 5 bona-fide~\footnote{Excluding the proximate  $z_{\rm qso}\simeq z_{\rm dla}$ DLAs.} 
intervening DLAs which have been identified spectroscopically in emission 
\citep{LowenthalJ_91a,DjorgovskiS_96a,MollerP_04a,HeinmullerJ_06a,FynboJ_10a}. 
Interestingly, the recent host discovered by \citet{FynboJ_10a} towards Q2222-09 meets our \MgII\ criterion with $\EW\sim2.7$\AA, and has a large SFR of $\sim17$\mpy\ (uncorreced for dust)
from recent SINFONI observations \citep{PerouxC_11c}.
There is also the one candidate of 
\citet{FumagalliM_10a} who used the `Lyman limit' technique of \citet{OMearaJ_06a} to image DLA hosts
behind a Lyman limit system used as blocking filter.~\footnote{The disadvantage of this technique, which relies on broad-band imaging, is that only candidates can be found.}
Recently, some have argued that a `metallicity' criterion such as $\SiII1526>1$\AA\  
\citep[e.g.][]{FynboJ_10a} may help in pre-selecting detectable hosts.
Evidence for such a metallicity boost is emerging from the survey of \citet{PerouxC_11c} where
they found, over a wider range of redshift,
 a 30\%\ success rate for high-metallicity DLAs (with $[\Zn/H]>-0.65$) versus 10\%\ for the low metallicty sample.
In other words,  the $z=2$ detection rate might still be low even for high-metallicity DLAs.

Overall, our low detection rate in conjunction with these literature results are at odds with recent cosmological simulations.
Using the state-of-the-art adaptive mesh-refinement  (AMR) code ENZO, \citet{CenR_10a} found that, in cosmological simulations,
 DLAs can arise in a wide variety of environments, from cold gas clouds in galactic
disks to cold streams to cooling gas from galactic winds to cool clouds entrained by hot
galactic winds. However, the prediction at $z=2$ is that DLAs occur within $<30$~kpc of a blue star-forming galaxy, with
a SFR of $0.3$--$30$~\mpy, peaking at 6-8~\mpy.  The cosmological simulation of \citet{CenR_10a}
predicts that about two thirds (or 6) of the hosts should meet our survey limits, whereas we detected \NdHI. 
\citet{PontzenA_08a} found that the majority of DLAs reside in halos
with $10^{10}$\msun\ and near galaxies with star formation rates with $>1$\mpy.
All these models show that DLAs typically arise a few kpc away from galaxies
that would be identified in emission with significant SFRs.  
Our low detection rate --in conjunction with the literature results-- 
 disfavors these models.

\section{Conclusions}

In summary, we used the SINFONI IFU to
 search for  the \Ha\ signature of the hosts towards \NobsMg\ \MgII\ absorbers with \EW$>2$~\AA.
We found the following.

$\bullet$ Only 3 hosts out of our sample of \NobsMg\ \MgII\ absorbers with \EW$\ga2$~\AA\
were detected (Fig.~\ref{fig:maps}) down to SFR$\ga\sfrmin$~\mpy\ ($5\sigma$).
The detected hosts are found  with impact parameters of $\sim15$--30~kpc, reside in haloes with $\Mv\sim2$--3$\times10^{11}$\msun,
 have dynamical masses $M_{\rm 1/2}\sim2\times10^{10}$\msun\ and SFRs ranging from 9 to 18~\mpy;

$\bullet$ Such a  low detection rate (\Nd/\Nmgii) ---including two other SINFONI results from \citet{PerouxC_11c}--- is in sharp contrast to our $z=1$ SIMPLE survey where we detected 14 out of 21 hosts \citep{BoucheN_07a}. 
Taking into account the difference in sensitivity between the two surveys, we should have been able to detect $\ga\NeEvol$ ($\ga$\Ne) of the \Nmgii\
 \MgII\ hosts assuming an evolution of the SFRs as $\propto(1+z)^{\gamma}$ with $\gamma=2.5$ ($\gamma=0$), whereas  only  \Nd\ were detected.  
This is statistically significant at the  \NsigmaEvol$\sigma$ (\Nsigma$\sigma$) confidence level, respectively;

$\bullet$ The SFR distribution  at $z=2$ is strikingly different to that at $z=1$ (Fig.~\ref{fig:SFRdistri}).  
At $z\simeq2$, all three detections have SFR$\ga9$~\mpy, while at $z=1$, all the hosts have SFR$\la9$~\mpy.
 The increase from $z=1$ to $z=2$ in the SFRs for the detected hosts  is consistent with the evolution of the  SFR--mass 
sequence, which goes as $\propto (1+z)^{2.7}$.

$\bullet$  The undetected $z=2$ hosts cannot be hidden under the quasar continuum. 
After stacking our SINFONI cubes, we find no detectable \Ha\ emission within 0.3\arcsec\ of the QSO to our SFR limit
(Fig.~\ref{fig:stacked}).
Furthermore, the hosts cannot just be outside our surveyed area of $10\arcsec\times10$\arcsec\ (80~kpc$\times$80~kpc).
This  would run against our sample selection criteria ($\EW>2$\AA) 
and the  known \EW--impact parameter relation for low-ionization ions \citep[e.g.][]{LanzettaK_90a,SteidelC_95a}.
For instance, the anti-correlation between  \SiII1526 rest-EW and $\rho$ of \citet{SteidelC_10a}
would place all of our hosts within $40$ kpc given that our sample has a mean rest-EW of 0.8\AA\ in \SiII1526.

It is thus very unlikely that they be outside the field of view
since this would contradict the recent observations of \citet{SteidelC_10a}.

$\bullet$ We also search for the host galaxy towards \NobsHI\ \HI-selected absorbers, and found only 1 \Ha\ emitter.
Including the 12 sight-lines from \citet{PerouxC_11c},  only \NdHI\ \HI-absorbers out of \Nhi\ sight-lines have been detected
with SINFONI.
This low detection rate (\NdHI/\Nhi) disfavors the interpretation of \citet{RafelskiM_11a} that DLAs are probing the outskirts of LBGs.

\vspace{0.5cm}

Our low-detection rate directly imply that the $z=2$ \MgII\ hosts
cannot reside  in haloes with similar masses than those at $z=1$ which have $\Mv\sim10^{11.3}$ \msun.
Otherwise their SFRs would have been
well above our limit.  Consequently, strong \MgII\ systems could reside 
(i) in haloes with very low masses  ($\Mv\la10^{10.6}$ \msun), or (ii) in haloes with intermediate
mass with $\Mv\sim10^{10.8}$ \msun.
The first option (option B in Fig.~\ref{fig:MyFig}) 
is consistent with the wind scenario, in which the large equivalent widths are caused by the cold ($T\sim10^4$~K)
 material entrained in the super-nova driven outflows generated in very low mass galaxies.  
Under this scenario, in order to not violate the observed
number of absorbers $dN/dX$, the cross-section of $\EW>2$\AA-selected winds ought to be $\leq20$~kpc.
However, this  wind scenario cannot explain the detections and non-detections at the same time since
the \EW-to-SFR ratio must have evolved in opposite directions from $z=1$ for the two sub-samples.

An alternative option (option A in Fig.~\ref{fig:MyFig})
is that $z=2$ \MgII\ absorbers  reside in haloes of intermediate mass with $\Mv\sim10^{10.6\hbox{--}11.0}$\msun.
In this case, $z=2$ galaxies traced by strong \MgII\ absorbers do not form stars at a rate expected for their halo mass,
 in contrast to LBGs (and all other SF galaxies) which do form stars at a rate expected for their halo mass
provided that the accretion efficiency $\epsilon=1.0$ is high \citep[e.g.][]{GenelS_08a}.  
Since we did not detect the majority of the galaxies, 
this scenario implies that the efficiency $\epsilon$ for accretion is  much smaller at those mass scales
otherwise the SFRs would have been   above our observational limit. 
In other words, this would support  the existence of a transition in accretion (and/or galaxy properties) at $\Mv\sim10^{11}$~\msun,
where cooling is less efficient in low mass haloes. 
This scenario implies that $\epsilon(\Mv)$ goes as $\Mv^{\beta}$ with $\beta>1$. 
A consequence of this steep mass-dependence
is that this scenario can explain both the detections and the non-detections by invoking
a finite scatter around a $\Mv$--\EW\ relation.

  Clearly, some star-formation must have occurred in the past to produce these metals.
The very nature of \MgII\ systems means that SF occurred and enriched these regions (with $\alpha$-elements) at an earlier epoch.
This pre-enrichment scenario is 
 supported by the recent analysis of the clustering of \CIV\ systems by \citet{MartinC_10a}, where
they found that the size of enriched regions  around $z=2$ to $z=4.5$ galaxies is large $\sim0.4$ Mpc (co-moving).
The implied time scale for dispersing metals to such distances is larger than the typical stellar ages of SF
galaxies. This  means that enrichment by (low-mass) galaxies at an earlier epoch $z>4.5$ must have occurred
to account for the metal enriched regions traced by strong \MgII\ intervening absorbers.

\section{Acknowledgments}
We thank Kyle Stewart and Shy Genel for discussions.
We thank the ESO staff for their support in carrying these observations
and  take the opportunity to
thank the entire SINFONI instrument team.  
We thank the anonymous referee for a constructive and detailed report that led to 
significant improvements of the draft.
This research was supported by a Marie Curie International Outgoing
Fellowship (PIOF-GA-2009-236012) within the 7th European Community Framework Programme.
MTM thanks the Australian Research Council for a QEII Research Fellowship (DP0877998).
We acknowledge the use of the Sloan Digital Sky Survey (SDSS). 
Funding for the SDSS and SDSSII
has been provided by the Alfred P. Sloan Foundation, the Participating
Institutions, the National Science Foundation, the U.S.
Department of Energy, the National Aeronautics and Space Administration,
the Japanese Monbukagakusho, the Max Planck Society,
and the Higher Education Funding Council for England.


\begin{table*}
\caption{Summary of new observations. \label{table:observations}}
\begin{tabular}{llllccccccc}
\hline
QSO	 		& $z_{\rm qso}$ & $z_{\rm abs}$ & $W_r$(\AA) 	& Ref. &  PSF($\arcsec$)  &  $t_{\rm exp}$(s) &   Run ID &  Dates     \\
(1)   		&	(2)	&	(3)	& (4)	&  (5)&  (6)  &  (7)	& (8) & (9)\\
\hline
\MgII-selected\\
\hline
SDSSJ031522.09-080043.7 & 2.8940 	& 2.06056 	& 2.9/2.9	&(1)&  0.7		& 3000	&	076.A-0527 	& 	 2005-10-07 	\\
SDSSJ091247.59-004717.3 & 2.8590 	& 2.07097 	& 2.3/2.2	&(1)&  0.6		& 2400	&	076.A-0527 	&	2006-03-18	\\
SDSSJ103446.54+110214.4 & 4.2660 	& 2.11605 	& 2.9/1.8	&(1)&  0.6  	&  4800	&	081.A-0682 &  	2008-03-31 \\
SDSSJ104252.32+011736.5 & 2.440		& 2.2667	& 2.3/2.2 	&(1)&  0.6 		& 4800 	&	081.A-0682 	&	2008-03-30 \\
SDSSJ104747.08+045638.6 & 2.1220	& 2.07129 	& 2.8/2.3	&(1)&  0.6		& 2400	&	076.A-0527 		&	2006-03-18 \\
SDSSJ111008.61+024458.0 & 4.1170 	& 2.11874 	& 2.6/2.9	&(1)&  0.6 		&  12000	& 	081.A-0682  		&  	2008-03-31 	 \\
	 & 		& 		& 		&	&		&  	& 		  082.A-0580 	& 	 2009-01-10 \\
SDSSJ114436.65+095904.9 & 3.1500 	& 2.09277	& 4.1/3.3	&(1)&  0.6		&  4800	&	081.A-0682& 	2008-04-02 \\
J1205-0742 		& 4.694		& 2.44400 	& 3.9/2.8	& (1) &  0.6		&  4800	&	081.A-0682	&	2008-04-21 \\
SDSSJ125525.67+030518.4 & 2.5300	& 2.11441 	& 3.2/2.9	&(1)&  0.5		&  4800	& 	082.A-0580 &  	2009-02-29 \\
SDSSJ130907.93+025432.6 & 2.9400 	& 2.24588	& 2.1/1.1	&(1)&  0.7		&  4800 &	082.A-0580 	& 	2009-01-22 	\\ 
	 & 	 	& 		&		& &  		&  	 &		& 	 2009-02-07\\ 		
SDSSJ131625.40+124411.8 & 3.0940 	& 2.03673 	& 3.4/2.0	&(1)&  0.7 		&  2400	&	079.A-0341  & 	2007-03-19 \\
SDSSJ132139.86-004151.9 & 3.0740	& 2.22157 	& 3.7/2.5	&(1)&  0.6		&  4800	&	081.A-0682 	& 	2008-04-23 \\
SDSSJ143500.27+035403.5 & 2.4920 	& 2.27065 	& 2.6/2.3  	&(1)&  0.9          &  2400	& 	079.A-0600& 	2007-04-19 	\\
SDSSJ151824.37-010149.8 & 2.5860 	& 2.03632 	& 2.0/2.1	&(1)&  0.5		&  4800	&	082.A-0580 	& 	2009-02-28 \\
SDSSJ161526.64+264813.7 & 2.1800 	& 2.11728 	& 4.4/4.3	&(1)&  0.6		&  4800	&	081.A-0682 	&	2008-04-02 \\
SDSSJ205724.14-003018.7 & 4.6630	& 2.26871 	& 2.2/2.2	&(1)&  0.7		&  4800 &	081.A-0682&	2008-10-04 \\
Q2243-60\footnotemark[1] 		& 3.01 		& 2.3288  	& 2.6	& (2)	&  0.75 	&  9600  &  060.A-9041  & 	 2004-08-17/19 \\  
SDSSJ233156.47-090802.0 & 2.6610 	& 2.14265 	& 1.9/2.6 	&(1)&  0.8 		&  2400	&  	076.A-0527  &	2005-10-08 	\\
\hline
\HI-selected  & & & $\log $\NHI & \\
\hline
SDSSJ004732.73+002111.3 & 2.8788	& 2.4687	& 20.0 & (3)	& 0.6		& 6000	& 	081.A-0682	&	2008-09-15\\
SDSSJ013317.79+144300.3 & 3.2323	& 2.4754 	& 20.0 & (3)	& 0.6		& 4800	&	081.A-0682	&	2008-09-15 \\
Q0216+080 		& 2.99 		& 2.2931	& 20.50  & (4)	&  0.8		&  9600		& 060.A-9041  &     2004-08-14/15 	\\
Q1037-27  	 	&  2.19 	& 2.13900 	&  19.70 & (5)	& 0.6		& 2400 	& 	079.A-0341 &  2007-04-18 \\
SDSSJ131757.98+055938.6	& 2.3111	& 2.1742	& 20.05 & (3)	& 0.6		&  1200	&	081.A-0682	&	2008-04-15 \\
SDSSJ143912.04+111740.5 & 2.5827	& 2.4184 	& 20.25  & (6)	&  1.0		&  2400	&	081.A-0568	&       2008-04-09 	\\
Q2243-60\footnotemark[1] 		
			& 3.01 		& 2.3288  	& 20.67 & (2)	&  0.75 	&  9600   	&  060.A-9041  & 	 2004-08-17/19 \\  	
\hline 
\end{tabular}					\\	   
{(1) QSO name;
(2) QSO emission redshift;
(3) Absorber redshift;
(4) \MgII\ rest-equivalent width ($\AA$) / $\log$ \NHI (cm$^{-2}$);
(5) References for $W_r$ or $\log$ \NHI 
(1: This work; 
2: From \citet{LopezS_02a}; 
3: From \citet{ProchaskaJ_05a}; 
4: From \citet{LedouxC_06a}
5: From \citet{RyabinkovA_03a};
6: From \citet{SrianandR_08a}
);
(6) FWHM of the seeing PSF;
(7) Exposure time;
(8) Observing run ID;
(9) Dates of observations.
}\\     
$^1${Source common to both samples.}
\end{table*}

\begin{table*}
\caption{Combined Sample Properties. \label{table:summary}}
\begin{tabular}{llllccccccccc}
\hline
Sight Line 		
& $z_{\rm qso}$ 
& $z_{\rm abs}$ 
& $W_r$(\AA) 	
& $\log $\NHI
& $5\sigma_{\Ha}$
& $5\sigma_{\rm SFR}$	
&  $\Delta_{RA},\Delta_{Dec}$	
&	$\rho$ 
& $f_{\Ha}$
& SFR & SFR$_0$ & Ref.  \\
	   (1)		&	(2)	&(3)			& (4)	& (5) 	&	 (6)	& (7)	&  (8) & (9)	& (10)	&  (11)	& (12)	 & (13) \\
\hline
\MgII-selected			&		&			&	&		&& $<\fluxmin$	& $<\sfrmin$  &\\
\hline
SDSSJ031522.09-080043.7 & 2.8940 	& 2.06056 	& 2.9/2.9	& 0		& $<2.8$ & $<4.0$  	&	$\cdots$ & $\cdots$  &	$\cdots$ & $\cdots$& $\cdots$& (1)\\
SDSSJ091247.59-004717.3 & 2.8590 	& 2.07097 	& 2.3/2.2	& 0		& $<4.0$ & $<7.3$ 	&	$\cdots$ & $\cdots$ &	$\cdots$ & $\cdots$ & $\cdots$& (1)\\
SDSSJ103446.54+110214.4 & 4.2660 	& 2.11605 	& 2.9/1.8	& 0	  	& $<1.3$ & $<2.0$  		&	$\cdots$  & $\cdots$ & 	$\cdots$ & $\cdots$ & $\cdots$& (1)\\
SDSSJ104252.32+011736.5 & 2.440 	& 2.2667	& 2.3/2.2	& $\cdots$	& $<1.0$ & $<1.7$ 	&	$\cdots$  & $\cdots$ & 	$\cdots$ & $\cdots$& $\cdots$& (1)\\
SDSSJ104747.08+045638.6 & 2.1220	& 2.07129 	& 2.8/2.3	& 0		& $<4.0$ & $<7.2$  &		$\cdots$ & $\cdots$ &	$\cdots$ & $\cdots$& $\cdots$ & (1)\\
SDSSJ111008.61+024458.0 & 4.1170 	& 2.11874 	& 2.6/2.9	& 1	 	& $<0.8$ & $<1.2$  & 	 	+2.0,-0.25  & 16.6  & 6.0  & 9.1  & 19.2	& (1)\\
SDSSJ114436.65+095904.9 & 3.1500 	& 2.09277	& 4.1/3.3	& 1	 	& $<1.2$ & $<1.5$ 	&	(+1.2,+1.85)  & (18.6)$^4$  & $\cdots$  & $\cdots$  & $\cdots$    & (1) \\
J1205-0742 		& 4.694		& 2.44400 	& 3.9/2.8	& 0	 			&  $<1.6$ & $<3.5$ 	& 	$\cdots$ & $\cdots$ &	$\cdots$ & $\cdots$ & $\cdots$& (1)\\
SDSSJ125525.67+030518.4 & 2.5300	& 2.11441 	& 3.2/2.9	& 1		&  $<0.9$ & $<1.4$ 	&	+1.125,-1.25 & 14.1	&   5.6  & 8.6 & 18.0	& (1)  \\
			&		&		&		&		&    &  		& 	+4.00,+2.37 &   38.7	&   7.1  & 10.8 & 22.6& (1)\\
SDSSJ130907.93+025432.6 & 2.9400 	& 2.24588	& 2.1/1.1	& 0		&  $<1.2$ & $<2.1$ & 		$\cdots$ & $\cdots$ &  	$\cdots$ & $\cdots$& $\cdots$ & (1)\\ 	
SDSSJ131625.40+124411.8	 & 3.0940 	& 2.03673 	& 3.4/2.0	& 0		&  $<4.0$ & $<5.6$    		&	$\cdots$ & $\cdots$&	$\cdots$ & $\cdots$& $\cdots$ & (1)\\
SDSSJ132139.86-004151.9 & 3.0740	& 2.22157 	& 3.7/2.5	& 0		&  $<1.2$ & $<2.1$  		&	$\cdots$ & $\cdots$ &	$\cdots$ & $\cdots$& $\cdots$& (1)\\
SDSSJ143500.27+035403.5 & 2.4920 	& 2.27065 	& 2.6/2.3  	& 1	  	&  $<1.8$ & $<3.0$   	&  	$\cdots$ & $\cdots$ 	&	$\cdots$ & $\cdots$& $\cdots$& (1)\\
SDSSJ151824.37-010149.8 & 2.5860 	& 2.03632 	& 2.0/2.1	& 1	 	&  $<1.2$ & $<1.7$  		& 	$\cdots$ & $\cdots$ &	$\cdots$ & $\cdots$& $\cdots$ & (1)\\
SDSSJ161526.64+264813.7 & 2.1800 	& 2.11728 	& 4.4/4.3	& 0	 	&  $<1.1$ & $<1.7$  		&	$\cdots$ & $\cdots$ & 	$\cdots$ & $\cdots$& $\cdots$& (1) \\
SDSSJ205724.14-003018.7 & 4.6630	& 2.26871 	& 2.2/2.2	& 0		&  $<1.2$ & $<2.2$  		&	$\cdots$ & $\cdots$ &	$\cdots$ & $\cdots$& $\cdots$ & (1)\\
Q2243-60 		& 3.01 		& 2.3288  	&  2.6$^{\footb}$  & 20.67  	&  $<0.6$ & $<1.3$   &	-2.1,-2.1 & 26.5 & 8.0 & 17 & 36 & (1)   \\ 
SDSSJ233156.47-090802.0 & 2.6610 	& 2.14265 	& 1.9/2.6 	& 1		&  $<2.4$ & $<3.6$ 		&	$\cdots$ & $\cdots$ &	$\cdots$ & $\cdots$& $\cdots$ & (1) \\
\hline
SDSSJ205922.42-052842.7$^{\foota}$   &  2.539 & 2.2100 & 2.1/1.7  	& 20.80  & $<1.6$ 	& $<2.7$ 	&	$\cdots$ & $\cdots$&	$\cdots$ & $\cdots$& $\cdots$ & (2) \\
SDSSJ222256.11-094636.2$^{\foota}$   &  2.927 & 2.3543 & 2.7$^{\footc}$  & 20.50   & $<1.1$ & $<2.2$ 	  & +0.5,+0.5 &   6 &  9.0 & 18 & 37 & (2) \\
\hline
\hline
\HI-selected			&		&		&		&		&	 &  $<\HIlim$	&		&\\
\hline
SDSSJ004732.73+002111.3 & 2.8788	& 2.4687	& $\cdots$ 	& 20.00		 &  $<2.1$ & $<4.9$ 		&  	$\cdots$ & $\cdots$    &	$\cdots$ & $\cdots$ & $\cdots$ & (1) \\
SDSSJ013317.79+144300.3 & 3.2323	& 2.4754 	& $\cdots$	& 20.00		  &  $<1.9$  & $<4.3$	& 	$\cdots$ & $\cdots$ &	$\cdots$ & $\cdots$ & $\cdots$ & (1) \\
Q0216+080		& 2.99 		& 2.29		& $\cdots$	& 20.50 	  &  $<1.0$  & $<1.8$		&  	$\cdots$ & $\cdots$ &	$\cdots$ & $\cdots$ & $\cdots$ & (1)\\
Q1037-27 	 	&  2.19 	& 2.13900 	& $\cdots$	& 19.70		 &  $<3.0$  & $<4.7$  	& $\cdots $ & $\cdots $&	$\cdots$ & $\cdots$ & $\cdots$ & (1)\\
SDSSJ131757.98+055938.6   & 2.3111	& 2.1742	& 1.6/1.3	& 20.05		& $<2.4$ & $<4.3$ 	 	&   	$\cdots$ & $\cdots$ &	$\cdots$ & $\cdots$ & $\cdots$ & (1)\\
SDSSJ143912.04+111740.5 & 2.5827	& 2.4184 	&  $\cdots$ 	& 20.25		&  $<2.3$ & $<3.9$		&  	$\cdots$ & $\cdots$ &	$\cdots$ & $\cdots$ & $\cdots$ & (1)\\
Q2243-60		& 3.01 		& 2.3288  	&  2.6$^{\footb}$		& 20.67  	&  $<0.8$ & $<1.5$    &	-2.1,-2.1 & 26.5 & 8.0 & 17 & 36   & (1) \\  
\hline   
SDSSJ121134.95+090220.8   &  3.292 & 2.5841 	& $\cdots$ 	& 21.40   	& $<5.6$ 	& $<14$			&	$\cdots$ & $\cdots$&	$\cdots$ & $\cdots$&	$\cdots$  &(2) \\
SDSSJ122607.19+173649.8   &  2.925 & 2.5576 	& $\cdots$ 	& 19.32   	& $<6.8$ 	& $<17$			&	$\cdots$ & $\cdots$&	$\cdots$ & $\cdots$&	$\cdots$  &(2) \\
Q1228-113  	 	& 3.528 	& 2.1929 		& $\cdots$  &  20.60  	& $<0.8$  	& $<1.4$ 	 & $\cdots$ & $\cdots$&	$\cdots$ & $\cdots$&	$\cdots$  &(2) \\ 	 
Q1232+07		&  2.570 	& 2.3376 		& $\cdots$ 	& 20.80   	& $<0.8$	&  $<1.6$		&	$\cdots$ & $\cdots$&	$\cdots$ & $\cdots$&	$\cdots$  &(2) \\
Q1354-11   		&  3.006 & 2.5009 			& $\cdots$ 	& 20.40   	& $<2.2$ 	& $<5.2$	&	$\cdots$ & $\cdots$&	$\cdots$ & $\cdots$&	$\cdots$  &(2) \\
SDSSJ145418.58+121053.8   &  3.256 & 2.2550 	& 1.0/0.9	 & 20.30   	& $<2.1$ 	& $<3.7$	&	$\cdots$ & $\cdots$&	$\cdots$ & $\cdots$&	$\cdots$  &(2) \\
SDSSJ205922.42-052842.7$^{\foota}$&  2.539 & 2.2100 	& 2.1/1.6  	& 20.80   & $<1.6$ 	& $<2.7$	&	$\cdots$ & $\cdots$&	$\cdots$ & $\cdots$&	$\cdots$  &(2) \\
Q2102-35 			& 3.090  	& 2.5070 		& $\cdots$	 	& 20.21   & $<3.1$	&  $<7.2$	&	$\cdots$ & $\cdots$ &	$\cdots$ & $\cdots$&	$\cdots$  &(2) \\
SDSSJ222256.11-094636.2$^{\foota}$ &  2.927 & 2.3543 & 2.7$^{\footc}$ & 20.50   & $<1.1$ & $<2.2$  	& +0.5,+0.5 &   6 & 9.0 & 18 & 37  &(2) \\
Q2311-37     	& 2.476  	& 2.1821 		& $\cdots$ 	& 20.48   	& $<1.5$	&	$<2.5$	&	$\cdots$ & $\cdots$ &	$\cdots$ & $\cdots$&	$\cdots$  &(2) \\
SDSSJ235057.87-005209.9 	&  3.023 	& 2.6147 		& $\cdots$ 	& 21.30  	& $<4.0$		&	$<10$		&	$\cdots$ & $\cdots$ &	$\cdots$ & $\cdots$&	$\cdots$  &(2) \\
Q2359-01        	& 2.810  	& 2.0950 		& $\cdots$ 	& 20.70  	& $<1.3$	&  $<2.0$	 	&	$\cdots$ & $\cdots$&	$\cdots$ & $\cdots$  & $\cdots $  &(2) \\
\hline 
\end{tabular}	
{(1) Name of QSO sight-line; 
(2) QSO redshift; 
(3) Absorber redshift; 
(4) rest-frame \EW/$W_r^{2803}$; 
(5) \HI\ column density or `0/1' indicating whether the system meets the \citet{RaoS_06a} criteria for being a DLA; 
(6) \Ha\ flux limit ($5\sigma$) in $10^{-17}$\flux; 
(7) SFR limit ($5\sigma$) in \mpy\ for a Chabrier IMF; 
(8) R.A., Decl. offsets in \arcsec; 
(9) impact parameter in kpc; 
(10) \Ha\ line flux in $10^{-17}$\flux; 
(11) observed SFR in \mpy (Chabrier); 
(12) intrinsic SFR in \mpy\ corrected for dust;
(13) References (1: This work, 2: From the survey of \citet{PerouxC_11c}).
}		\\			        
$^{\foota}${Source common to both the \MgII\ and \HI\ sample};
$^{\footb}${From \citet{LopezS_02a}};
$^{\footc}${From \citet{FynboJ_10a}};
$^4${Continuum of a source detected with no redshift identification.}
\end{table*}

\begin{table*}

\caption{Sample Results\label{table:results}}
\begin{tabular}{lccccccccc}
\hline
Sample 	&  $N$	& SFRlimit	& $\hat p$		&	$N_e$	& $N_d$	& $P$-value($N_d$) & SFR$(z)$ \\
(1)	&  (2)	&	(3)	& (4)		&	(5)	&(6)		& (7) 	& (8)\\
\hline
\MgII	&  \Nmgii	&  \sfrmin	& 8/21		&   \Ne\		&  \Nd\		& \PvalMgii (\Nsigma$\sigma$)  & no Evol. \\
\MgII	&  \Nmgii	&  \sfrmin	& 12/21		&   \NeEvol\		&  \Nd\		& \PvalMgiiEvol\ (\NsigmaEvol$\sigma$) & $\propto(1+z)^{2.5}$ \\
\hline
\end{tabular}
\\
{(1) Sample name; (2) Number of sight-lines; (3) SFR limit ($3\sigma$) in \mpy; 
(4) Success rate expected from the z1SIMPLE results and corrected for the SFR limit;
(5) Number of detections expected;
(6) Number of actual detections;
(7) Probability that exactly $N_d$-detections occur by chance given the expected success rate $\hat p$;
(8) Assumed evolution of SFR($z$).
}
\end{table*}

\bsp

\label{lastpage}

\end{document}